\documentclass[a4paper]{article}

\usepackage{INTERSPEECH2022}

\title{Deep Learning for Speaker Identification: Architectural Insights from AB-1 Corpus Analysis and Performance Evaluation}
\name{Matthias Bartolo$^1$}
\address{
  $^1$Department of AI, University of Malta
  }
\email{matthias.bartolo.21@um.edu.mt}

\begin{document}

\maketitle
\begin{abstract}
\textbf{In the fields of security systems, forensic investigations, and personalized services, the importance of speech as a fundamental human input outweighs text-based interactions. This research delves deeply into the complex field of Speaker Identification (SID), examining its essential components and emphasising Mel Spectrogram and Mel Frequency Cepstral Coefficients (MFCC) for feature extraction. Moreover, this study evaluates six slightly distinct model architectures using extensive analysis to evaluate their performance, with hyperparameter tuning applied to the best-performing model. This work performs a linguistic analysis to verify accent and gender accuracy, in addition to bias evaluation within the AB-1 Corpus dataset.}
\end{abstract}

\section{Introduction}

Speaker identification (SID) is the task of determining a speaker's identity from a specific audio sample chosen from a pool of known speakers. With applications in forensics, security, and customization \cite{deeplearnsid}, SID may be expressed as a pattern recognition problem. The SID pipeline, according to \cite{ANTONY2018250}, is dependent on two critical components: \textit{feature extraction} and \textit{feature classification}. These factors work together to classify an input speech segment as belonging to one of N known enrolled speakers.

In feature extraction, certain characteristics required for an individual's identification are taken from the voice data. During the feature classification process, features extracted from an unidentified individual are compared to those of various speakers in order to detect and match the unique qualities that will eventually lead to the identification of the specific speaker \cite{deeplearnsid}. A perfect speaker identification system should have minimal intra- and inter-speaker variance, readily quantifiable attributes, be less susceptible to noise, respond correctly to imitation, and not rely on other qualities \cite{ANTONY2018250}.

\section{Feature Extraction}

\begin{figure}[ht]
  \centering
  \begin{minipage}[t]{0.4\textwidth}
    \centering
    \includegraphics[width=1\textwidth]{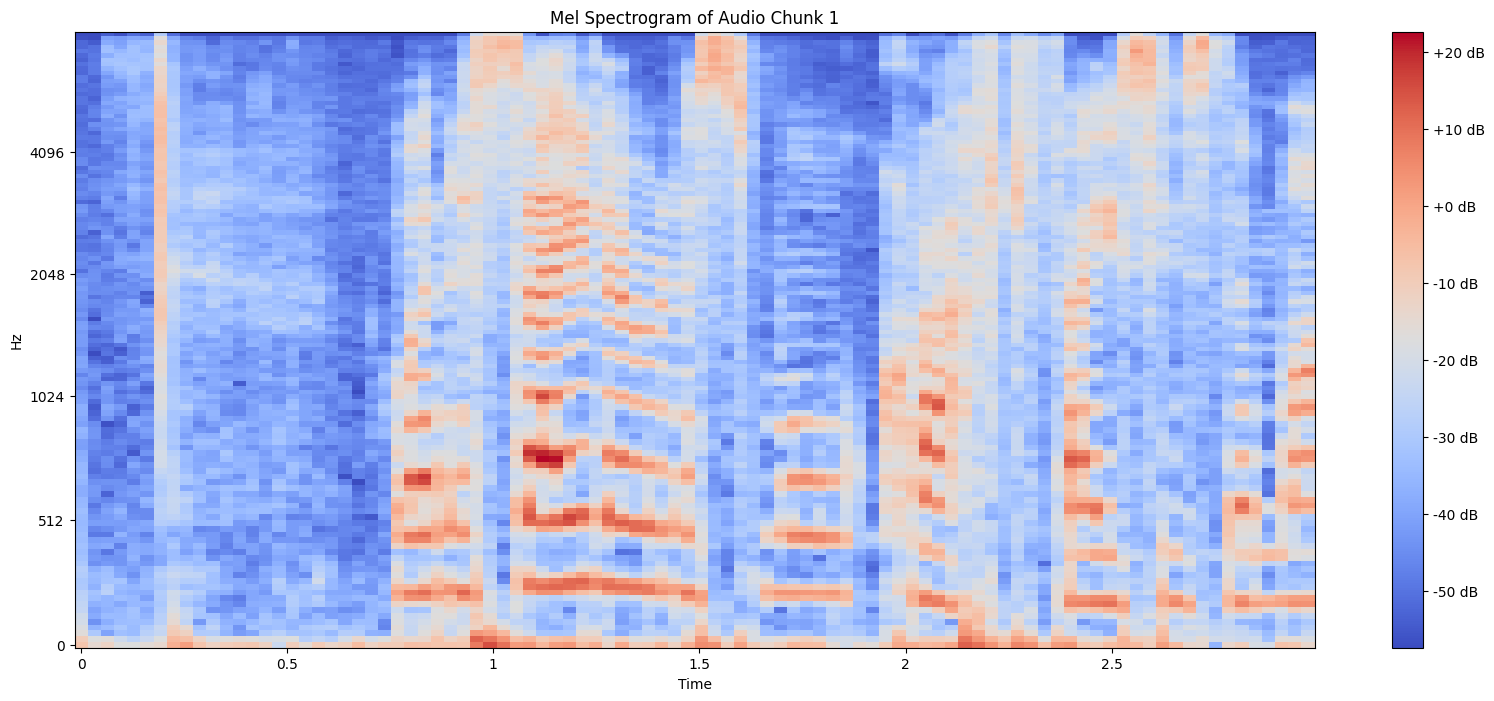}
    \caption*{Mel Spectrogram Feature Extraction}
    \label{fig:mel_spectrogram}
  \end{minipage}\hfill\hfill
  \begin{minipage}[t]{0.4\textwidth}
    \centering
    \includegraphics[width=1\textwidth]{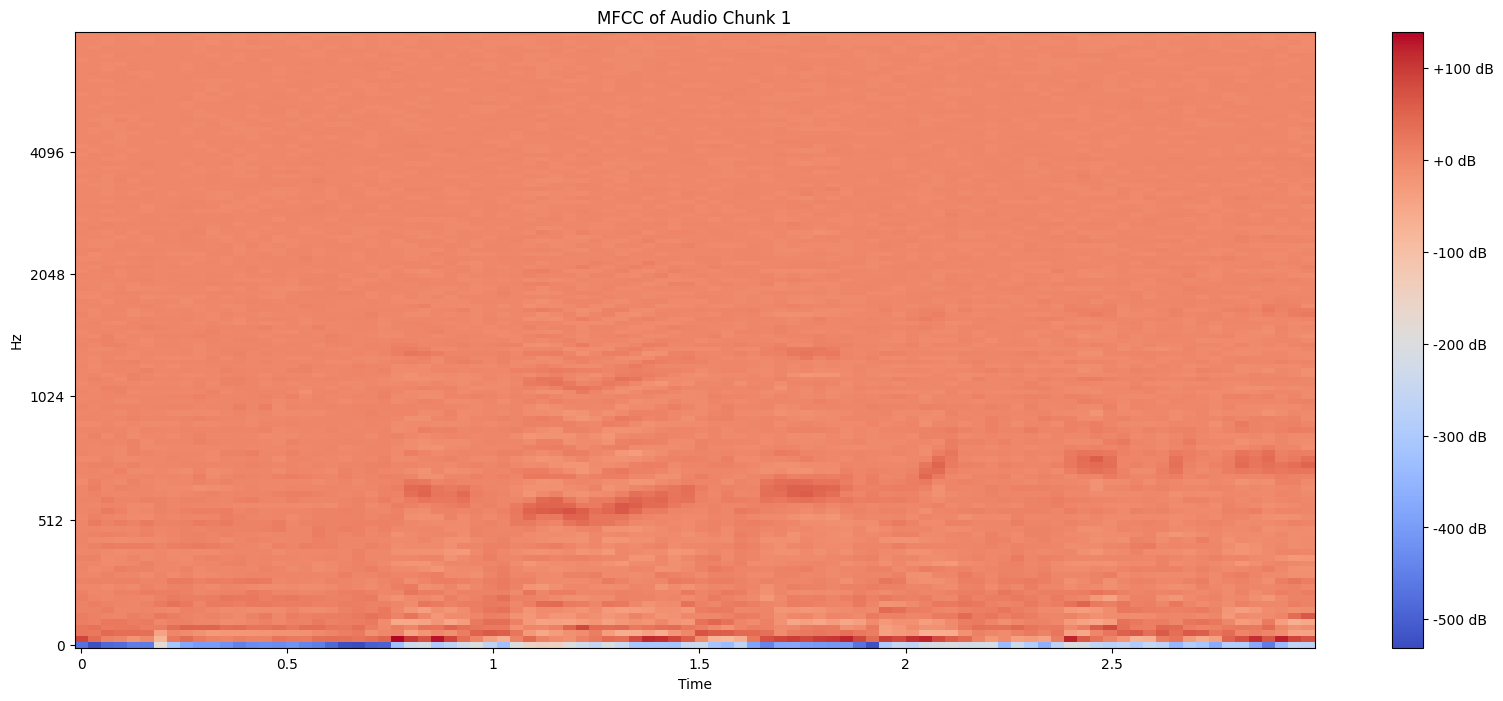}
    \caption*{MFCC Feature Extraction}
    \label{fig:mfcc}
  \end{minipage}
  \vspace{8pt}
  \caption{Comparison of different audio feature extraction methods for the same filtered three-second audio chunk.}
  \label{fig:audio_features_comparison}
\end{figure}

Feature extraction is a critical step in speech analysis, stemming from the fact that it allows raw audio data to be transformed into useful features. To this extent, \textit{Mel Spectrogram} and \textit{Mel Frequency Cepstral Coefficients (MFCC)} are two widely used techniques in this field of study \cite{deeplearnsid}. By translating frequencies into the Mel scale, the Mel Spectrogram visualises the frequency content of an audio source across time, emphasising human auditory perception. Meanwhile, MFCCs extract compact representations by capturing the spectral features of the audio signal \cite{ANTONY2018250}. In this study, both of the aforementioned feature extractors were tested as inputs to the developed system architectures, with the goal of producing a thorough analysis to evaluate their performance and appropriateness for the current context. Figure \ref{fig:audio_features_comparison} depicts the output of both extractors.



\section{Model Architecture}

\begin{table}[ht]
\centering
\small
\begin{tabular}{|c|l|l|}
\hline
\textbf{No.} & \textbf{Layer Type} & \textbf{Details} \\
\hline\hline
1 & Conv2D & (3, 3), 32 filters\\\hline
2  & ReLU & --\\
\hline
3 & Conv2D & (3, 3), 64 filters\\\hline
4  & ReLU & --\\
\hline
5 & MaxPooling2D & (2, 2)\\
\hline
6 & Conv2D & (3, 3), 64 filters\\\hline
7  & ReLU & --\\
\hline
8 & MaxPooling2D & (2, 2)\\
\hline
9 & Reshape & --\\
\hline
10 & LSTM & 64 units \\
\hline
11 & Flatten & --\\
\hline
12 & BatchNormalization & --\\
\hline
13 & Dropout & 30\% \\
\hline
14 & Dense & 285 neurons\\
\hline
15 & Softmax & 285 outputs \\\hline
\end{tabular}
\caption{Model 1 Architecture}
\label{tab:model_architecture}
\end{table}

This study investigated a total of six slightly distinct model architectures.
The first model architecture, as shown in Table \ref{tab:model_architecture}  incorporates elements from both architectures listed in \cite{text_independent_recognition} and \cite{lighten_cnn_lstm}. The model incorporates early feature extraction convolution layers modified from the structure suggested in \cite{text_independent_recognition}, followed by an altered single lightweight LSTM layer suggested by \cite{lighten_cnn_lstm}. The addition of batch normalization and dropout layers was intended to regularize the flattened LSTM output, hence improving the model's robustness and avoiding overfitting. Finally, classification was performed by activating softmax in the final fully connected dense layer. This function generated probability scores for each of the 285 speakers in the AB-1 corpus dataset, indicating the likelihood that the input data corresponds to one of the aforementioned speakers.

The succeeding model architectures in this research are all derived from the above architecture. Furthermore, the second architecture focused on raising the convolutional depth in the first feature extractors (CNNs) to improve feature extraction capabilities. Meanwhile, in the third model architecture, the LSTM was modified by increasing the number of LSTM units and incorporating an extra layer to increase sequence comprehension. The fourth architecture included an extra dense layer following the CNN-LSTM configuration to further enhance the output sequence. In contrast to the second model architecture, the fifth architecture was intended to reduce model complexity by recommending fewer convolution filters. Finally, the sixth architecture used batch normalisation across all CNN blocks in the model with the aim of reducing the effects of overfitting.

\section{Evaluation}

\begin{table*}[!htbp]
    \centering
    \resizebox{\textwidth}{!}{%
    \begin{tabular}{|c|c|c|c|c|c|c|c|c|}\hline
        \textbf{Model} & \textbf{Best Val Accuracy} & \textbf{Test Accuracy} & \textbf{Test Loss} & \textbf{Precision} & \textbf{Recall} & \textbf{F1-Score} & \textbf{Epoch Converged} & \textbf{Time Taken (minutes)}\\ \hline \hline
        {\textbf{\small Model-1-Mel}} & {\textbf{\small 0.97}} & {\textbf{\small 0.968}} & {\small 0.148} & {\textbf{\small 0.971}} & {\textbf{\small 0.968}} & {\textbf{\small 0.968}} & {\small 22} & {\small 30} \\ \hline 
        {\textbf{\small Model-1-MFCC}} & {\small 0.93} & {\small 0.928} & {\small 0.301} & {\small 0.937} & {\small 0.928} & {\small 0.928} & {\textbf{\small 13}} & {\small 20} \\ \hline 
        {\textbf{\small Model-2-Mel}} & {\small 0.01} & {\small 0.009} & {\small 5.619} & {\small 0.000}& {\small 0.009} & {\small 0.000}& {\small 19} & {\small 58} \\ \hline
        {\textbf{\small Model-2-MFCC}} & {\small 0.95} & {\small 0.928} & {\small 0.301} & {\small 0.937} & {\small 0.928} & {\small 0.928} & {\small 25} & {\small 78} \\ \hline
        {\textbf{\small Model-3-Mel}} & {\small 0.93} & {\small 0.934} & {\small 0.262} & {\small 0.947} & {\small 0.934} & {\small 0.934} & {\textbf{\small 13}} & {\small 18} \\ \hline
        {\textbf{\small Model-3-MFCC}} & {\small 0.91} & {\small 0.917} & {\small 0.343} & {\small 0.930}& {\small 0.917} & {\small 0.918} & {\small 21} & {\small 42} \\ \hline
        {\textbf{\small Model-4-Mel}} & {\small 0.96} & {\small 0.952} & {\small 0.252} & {\small 0.958} & {\small 0.952} & {\small 0.952} & {\small 24} & {\small 28} \\ \hline
        {\textbf{\small Model-4-MFCC}} & {\small 0.86}& {\small 0.841}& {\small 0.640}& {\small 0.862}& {\small 0.841}& {\small 0.838}& {\small 25}& {\small 31}\\ \hline
        {\textbf{\small Model-5-Mel}} & {\textbf{\small 0.97}} & {\small 0.965} & {\textbf{\small 0.141}} & {\small 0.968} & {\small 0.965} & {\small 0.964} & {\small 21} & {\small 13} \\ \hline
        {\textbf{\small Model-5-MFCC}} & {\small 0.93}& {\small 0.920}& {\small 0.313}& {\small 0.929}& {\small 0.920}& {\small 0.919}& {\small 19}& {\textbf{\small 12}}\\ \hline
        {\textbf{\small Model-6-Mel}} & {\small 0.96} & {\small 0.965} & {\textbf{\small 0.141}} & {\small 0.968} & {\small 0.965} & {\small 0.964} & {\small 19} & {\small 28} \\ \hline
        {\textbf{\small Model-6-MFCC}} & {\small 0.91}& {\small 0.920}& {\small 0.313}& {\small 0.929}& {\small 0.920}& {\small 0.919}& {\small 25}& {\small 38}\\ \hline \hline
        {\textbf{\small Best Model}} & {\textbf{\small 0.97}} & {\textbf{\small 0.971}} & {\textbf{\small 0.136}} & {\textbf{\small 0.974}} & {\textbf{\small 0.971}} & {\textbf{\small 0.971}} & {\textbf{\small N/A }}& {\textbf{\small N/A}}\\ \hline
    \end{tabular}%
    }
    \caption{Table of Results}
    \label{tab:results}
\end{table*}
The aforementioned model architectures were trained with the \textit{TensorFlow Keras framework}, which was chosen for its ease of use, efficacy in creating neural network models, and facilitation of efficient training methods. During training, the Adam optimizer, which is noted for its versatility and extensive use, was used \cite{kingma2017adam}. Additionally, an early stopping callback with a patience score of 5 was also included to prevent overfitting. 

\begin{figure}[ht]
    \centering
    \includegraphics[width=0.9\linewidth]{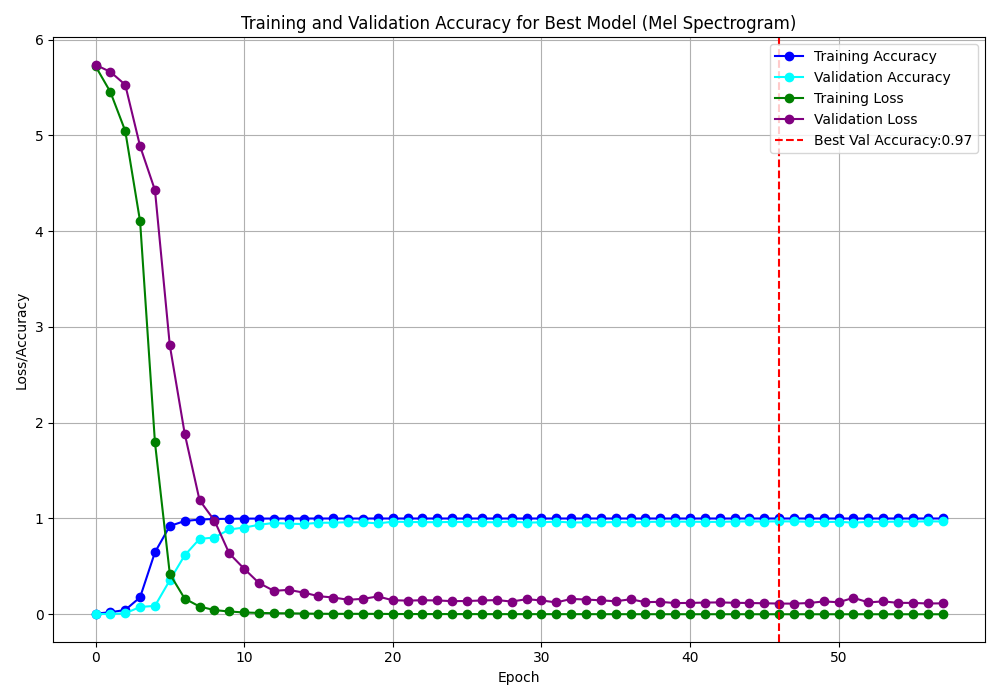}
    \caption{Curves demonstrating the best model's training and validation loss, as well as accuracy.}
    \label{fig:bet_model_curves}
\end{figure}

The findings shown in Table \ref{tab:results} reveal that models 1 and 5 outperformed all others in terms of test accuracy, precision, recall, and f-score. Overall, all the presented architectures worked relatively well, with the exception of model 2 using the MFCC feature extractor, which consistently yielded comparable poor results over several runs. Furthermore, the results also illustrate that models using the Mel Spectrogram feature extractor outperformed those using the MFCC feature extractor. Additionally, the models consistently produced test accuracy, precision, recall, and f-score metrics in the 0.8 to 0.97 range, indicating a noteworthy degree of performance. Furthermore, the test loss observed among the models ranged between 0.14 and 0.65, indicating that the models were not overfitting.

Following the discovery of model 1's superior efficiency, a thorough hyperparameter tuning procedure comprising fifteen trials was carried out to fine-tune its parameters.  From the test, the best hyperparameters found were a learning rate of \textbf{0.001}, a dropout rate of \textbf{0.4}, and the activation functions \textbf{tanh} applied specifically to the second layer in the model architecture presented in Table \ref{tab:model_architecture}, whilst the remaining layers utilised \textbf{relu}. The findings shown in Table \ref{tab:results} demonstrate a significant performance improvement attained by the best model refined with these optimised parameters over the original model 1.

The visible improvements in the curves in Figure \ref{fig:bet_model_curves}, which reflect the training and validation loss as well as accuracy for the best model, illustrate the efficiency of the adjusted parameters in optimising the model's performance. Furthermore, the curves exhibit a consistent and continuous evolution, demonstrating the model's incremental improvement during the training period, which is characterised by a decreasing loss and an increasing accuracy.

\begin{figure}[ht]
    \centering
    \includegraphics[width=0.9\linewidth]{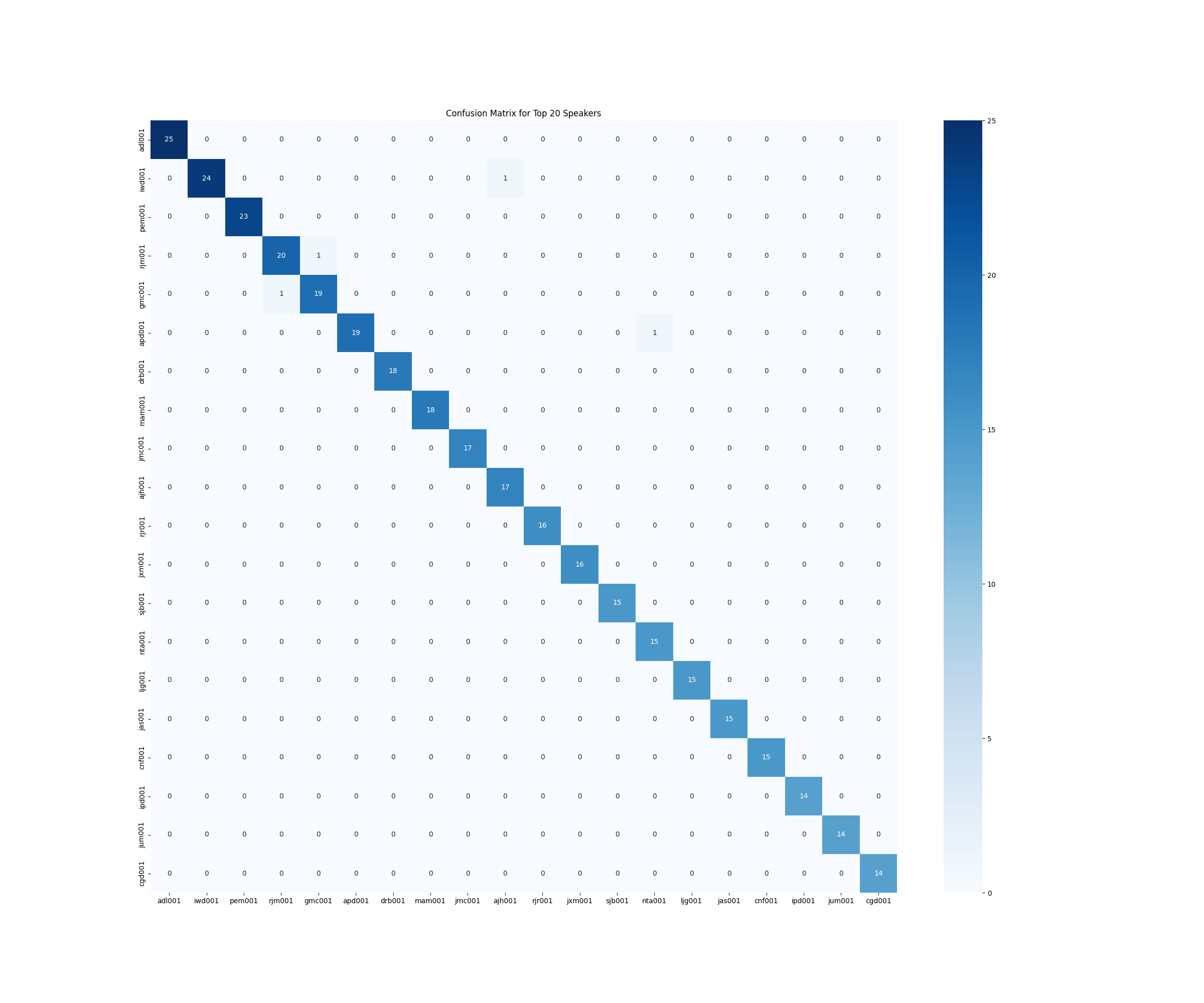}
    \caption{Confusion matrix showcasing the top 20 projected speakers for the best model.}
    \label{fig:confusion_matrix}
\end{figure}

The confusion matrix for the top 20 projected speakers in Figure \ref{fig:confusion_matrix} exhibits distinct diagonal components. Furthermore, these diagonal components, which particularly demonstrate the model's capacity to precisely anticipate speakers, are important indicators of the classification performance of the developed model.

\section{Analysis}
\begin{figure}[ht]
    \centering
    \includegraphics[width=0.9\linewidth]{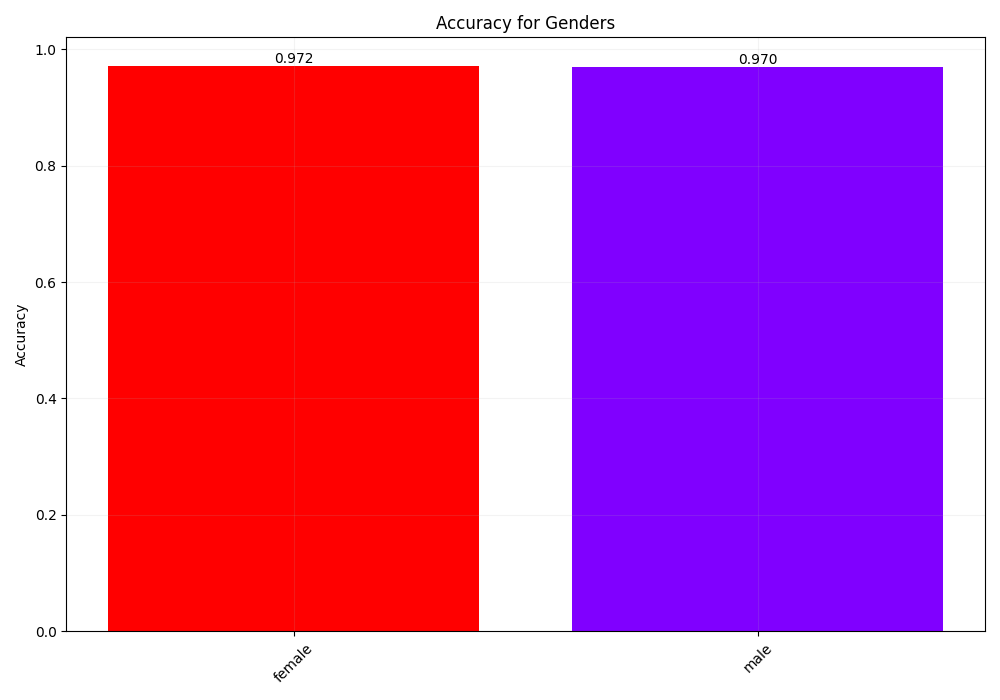}
    \caption{Gender accuracy and bias evaluation for the best model's performance on the test set.}
    \label{fig:gender_accuracy}
\end{figure}

A linguistic analysis was performed on the top-performing model to assess its gender and accent correctness while assuring the model's neutrality and lack of bias towards certain accents or genders. Figure \ref{fig:gender_accuracy} demonstrated a slight variance in gender accuracy. Surprisingly, the model performed somewhat better in predicting female speakers than male speakers, by a margin of 0.02. Furthermore, these findings imply that the model's predictions exhibit gender equality, indicating that there is equal representation within the AB-1 corpus dataset. This equality in predicted accuracy indicates that there is no bias towards any one gender, emphasizing a fair and impartial training dataset.

\begin{figure}[ht]
    \centering
    \includegraphics[width=0.9\linewidth]{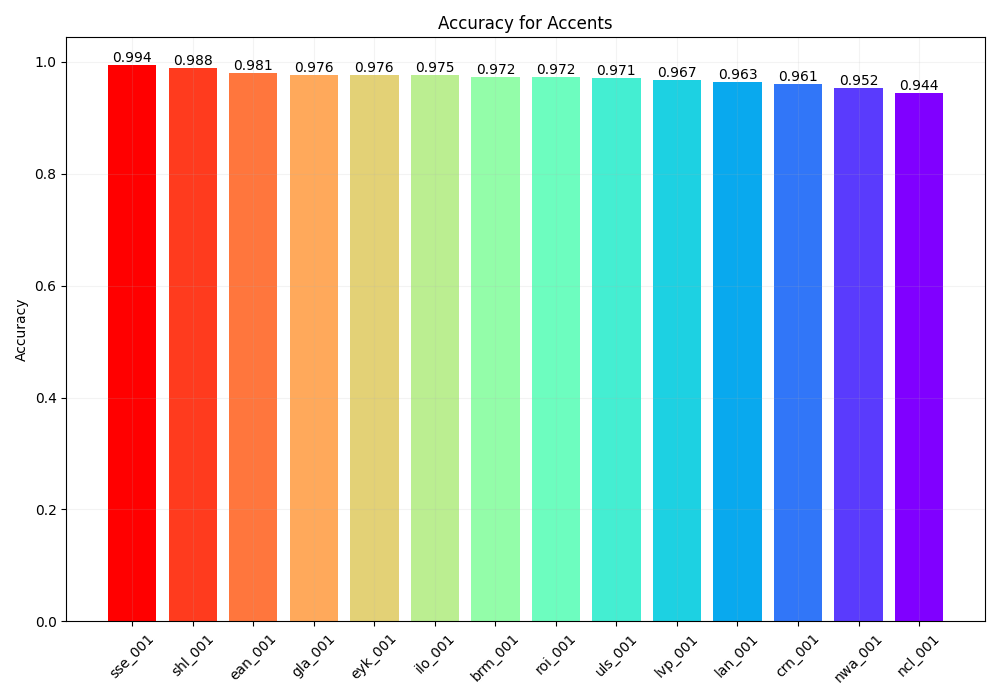}
    \caption{Accent accuracy and bias evaluation for the best model's performance on the test set.}
    \label{fig:accents_accuracy}
\end{figure}

As observed in Figure \ref{fig:accents_accuracy}, the accent accuracy scores were quite similar across accents. However, when compared to the gender accuracy distribution, a more obvious disparity between the best and lowest accuracy, approximately 0.05, emerged. From Figure \ref{fig:accents_accuracy} it can be deduced that the three easiest accents to predict (in order) were the following:
\begin{enumerate}
    \item Standard Southern English
    \item Scottish Highlands 
    \item East Anglia 
\end{enumerate}
In contrast, the three hardest accents to predict (in order) were the following:
\begin{enumerate}
    \item Newcastle
    \item Northern Wales
    \item Cornwall
\end{enumerate}

\section{Conclusion}
This paper investigates the efficacy of Mel Spectrogram and MFCC as feature extraction approaches for speaker identification (SID) while proposing robust model architectures designed particularly for SID. The study's best model accuracy, precision, recall, and F-score of 0.97 demonstrate the usefulness of these techniques. Gender accuracy analysis revealed minimal variation, with slightly greater accuracy in identifying female speakers, confirming dataset balance and the lack of gender bias in predictions. Nonetheless, accent accuracy resulted in a more noticeable discrepancy, with Standard Southern English being the most predicted and Newcastle being the least predictable. The results of this study highlight the significance of future analysis and model development, notably in the case of accent-related challenges in SID.

\bibliographystyle{IEEEtran}
\bibliography{references}

\end{document}